\documentstyle[preprint,prl,aps,epsfig]{revtex}
\tightenlines

\begin{document}

%%%%%%%%%%%%%%%%%%%%%%%%%%%%%%%%begin text%%%%%%%%%%%%%%%%%%%%%%%%%%%%%%%%%%%%
\preprint{\vbox{\hbox{BNL-67631}
\hbox{KEK Preprint 2000-68}
\hbox{PRINCETON/HEP/2000-5}
\hbox{TRI-PP-00-54}}}

%==============================================================================
\def\k{\(K^+\)}
\def\g{\(\gamma\)}
\def\cpi{\(\pi^+\)}
\def\cmu{\(\mu^+\)}
\def\npi{\(\pi^0\)}
\def\pisc{\(\pi\)scat}
\def\piscs{\(\pi\)scat's}
\def\Ke4{\( K^+\rightarrow \pi^+\pi^-e^+\nu_e\)}
\def\kppnn{\( K^+ \rightarrow \pi^+ \pi^0 \nu \bar\nu\)}
\def\bppnn{\( B(K^+ \rightarrow \pi^+ \pi^0 \nu \bar\nu)\)}
\def\ppnn{\( K_{\pi\pi\nu\bar\nu}\)}
\def\kpnn{\( K^+ \rightarrow \pi^+ \nu \bar\nu\)}
\def\pnn{\( K_{\pi\nu\bar\nu}\)}
\def\kpitwo{\( K^+ \rightarrow \pi^+\pi^0\)}
\def\pitwo{\( K_{\pi2}\)}
\def\kpitwg{\( K^+ \rightarrow \pi^+\pi^0\gamma\)}
\def\pitwg{\( K_{\pi2\gamma}\)}
\def\kmutwo{\( K^+ \rightarrow \mu^+\nu_{\mu}\)}
\def\mutwo{\( K_{\mu2}\)}
\def\kmuthr{\( K^+ \rightarrow \mu^+\pi^0\nu_\mu\)}
\def\muthr{\( K_{\mu3}\)}
\def\P{\(P_{\pi^+}\)}
\def\R{\(R_{\pi^+}\)}
\def\T{\(T_{\pi^+}\)}
\def\Eg{\(E_\gamma\)}
\def\Enpi{\(E_{\pi^0}\)}
\def\pimue{\(\pi^+\rightarrow\mu^+\rightarrow e^+\)}
\def\pimu{\(\pi^+\rightarrow\mu^+\)}
\def\Probnpi{\(Prob(\chi^2_{\pi^0})\)}
\def\Ng{\(N_\gamma\)}
\def\z{$z$}
%==============================================================================
\title{Search for the Decay $K^+ \to \pi^+ \pi^0 \nu\bar\nu$}
%%%%%%%%%%%%%%%%%%%%%%begin author list%%%%%%%%%%%%%%%%%%%%%%%%%%%

\author{
S.~Adler$^1$, M.~Aoki$^5$\cite{masa}, M.~Ardebili$^4$, M.S.~Atiya$^1$,
P.C.~Bergbusch$^{5,7}$, E.W.~Blackmore$^5$, D.A.~Bryman$^{5,7}$,
I-H.~Chiang$^1$, M.R.~Convery$^4$, M.V.~Diwan$^1$, J.S.~Frank$^1$,
J.S.~Haggerty$^1$, T.~Inagaki$^2$, M.M.~Ito$^4$, S.~Kabe$^2$,
S.H.~Kettell$^1$, Y.~Kishi$^3$, P.~Kitching$^6$, M.~Kobayashi$^2$,
T.K.~Komatsubara$^2$, A.~Konaka$^5$, Y.~ Kuno$^2$, M.~Kuriki$^2$,
T.F.~Kycia$^1$\cite{ted}, K.K.~Li$^1$, L.S.~Littenberg$^1$,
J.A.~Macdonald$^5$, R.A.~McPherson$^4$,
P.D.~Meyers$^4$, J.~Mildenberger$^5$, N.~Muramatsu$^2$, T.~Nakano$^3$,
C.~Ng$^1$\cite{susb}, T.~Numao$^5$, J.-M.~Poutissou$^5$, 
R.~Poutissou$^5$, G.~Redlinger$^5$\cite{george}, 
T.~Sato$^2$, T.~Shinkawa$^2$\cite{shink}, F.C.~Shoemaker$^4$, 
R.~Soluk$^6$, J.R.~Stone$^4$, R.C.~Strand$^1$,
S.~Sugimoto$^2$, C.~Witzig$^1$, and Y.~Yoshimura$^2$,
\\ (E787 Collaboration) }

\address{ 
$^1$Brookhaven National Laboratory, Upton, New York 11973\\
$^2$High Energy Accelerator Research Organization (KEK), 
Oho, Tsukuba, Ibaraki 305-0801, Japan \\ 
$^3$RCNP, Osaka University, 10--1 Mihogaoka, Ibaraki, Osaka 567-0047, 
Japan \\ 
$^4$Joseph Henry Laboratories, Princeton University, Princeton, 
New Jersey 08544 \\
$^5$ TRIUMF, 4004 Wesbrook Mall, Vancouver, British Columbia,
Canada, V6T 2A3\\
$^6$ Centre for Subatomic Research, University of Alberta, Edmonton,
Alberta, Canada, T6G 2N5\\
$^7$ Department of Physics and Astronomy, University of British Columbia, 
Vancouver, BC, Canada, V6T 1Z1
}

%%%%%%%%%%%%%%%%%%%%%%end author list%%%%%%%%%%%%%%%%%%%%%%%%%%%

\maketitle

\begin{abstract}
  The first search for the decay \kppnn\ has been performed with the 
  E787 detector at BNL.  Based on zero events observed in the kinematical 
  search region defined by $90\,$MeV/$c$ $<P_{\pi^+}<188\,$MeV/$c$ and 
  $135\,$MeV $<E_{\pi^0}<180\,$ MeV, an upper limit 
  \bppnn\ $<\ 4.3\ \times\ 10^{-5}$ at 90\%   confidence level
  is established.
\end{abstract}

\draft
%-----------------------------------------------------------------------------
\pagebreak
The  flavor-changing neutral-current
decay \kppnn\ (\ppnn) is interesting because of its
sensitivity to the combination of CKM mixing matrix 
elements $V_{ts}^{\ast}V_{td}$.  Knowledge of the branching ratio, \bppnn\,, 
would provide additional constraints on quark mixing parameters, 
complementary to and independent of limits set by the related process \kpnn.
Within the framework of the Standard Model (SM), \ppnn\ is dominated by 
short-distance contributions arising from one-loop penguin and box diagrams for
\(s\rightarrow d\nu\bar\nu\) mediated by charm and top 
quarks \cite{qcd}\cite{longd}.
Relating the hadronic matrix element of this decay to the measured 
isospin-rotated process \Ke4\ \cite{ke4} minimizes hadronic uncertainties, 
making possible a theoretically clean prediction.
However \ppnn\ is
highly suppressed by the GIM mechanism \cite{gim}, as well as by phase space,
so the SM prediction
for the branching ratio is small, \bppnn\ $=  (1-2) \times$ $10^{-14}$ \cite{ppnnbr}.
Because the present potential sensitivity for observing \ppnn\ falls many 
orders of magnitude short of this predicted level, the primary motivation is 
the search 
for physics beyond the SM.  Unlike \kpnn, \ppnn\ is sensitive to new physics
mediated by axial-vector as well as by vector currents.

Because the $\nu$ and $\bar\nu$ are not directly 
observable, a full event reconstruction of \ppnn\ is
not possible.  The signature is a single \cpi\ track, 
accompanied by two photons originating from the \npi, with
no other activity in the detector.  
A thorough accounting of all processes that can emulate the \ppnn\ event 
topology and demonstration of the means to reject them allow sensitivity
for observing \ppnn\ decays.

E787 is a rare-kaon decay experiment at the Alternating Gradient Synchrotron 
(AGS) facility at Brookhaven National Laboratory (BNL).  A  prolific source of
kaons is obtained by directing an intense beam of 24-GeV protons (typically
\(16\times10^{12}\) per 1.6-second AGS spill) onto a 6-cm long platinum 
production target.  A secondary beam containing $7 \times 10^6 ~K^+$ 
per spill is extracted at a momentum of 790 MeV/$c$ 
\cite{lesb3}.  Contaminants (i.e. \cpi\ and protons) are reduced by two 
stages of electrostatic separation to about $25\%$ of the beam at the
detector. Prior to entering the detector, beam particles are tracked 
and identified by a \v Cerenkov counter, two sets of multi-wire 
proportional chambers and a scintillation-counter hodoscope.  They also 
pass through a BeO degrader that slows the kaons to $300\,$MeV/$c$.
Approximately $20\%$ of the incident
kaons emerge from the degrader and enter the detector's stopping target.

\begin{figure}
\centerline{
\hfill
\psfig{file=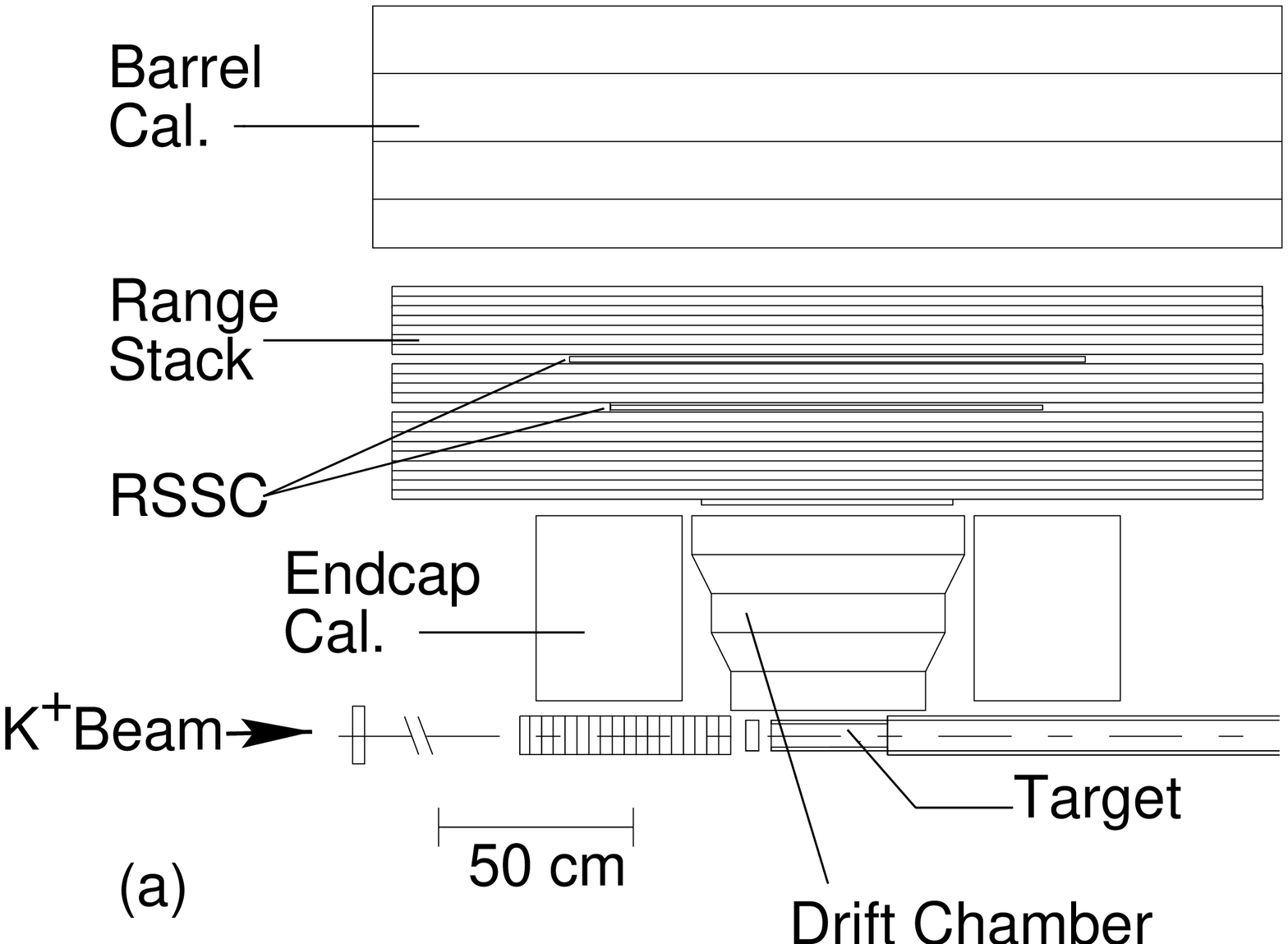,height=1.5in}
\hfill
\psfig{file=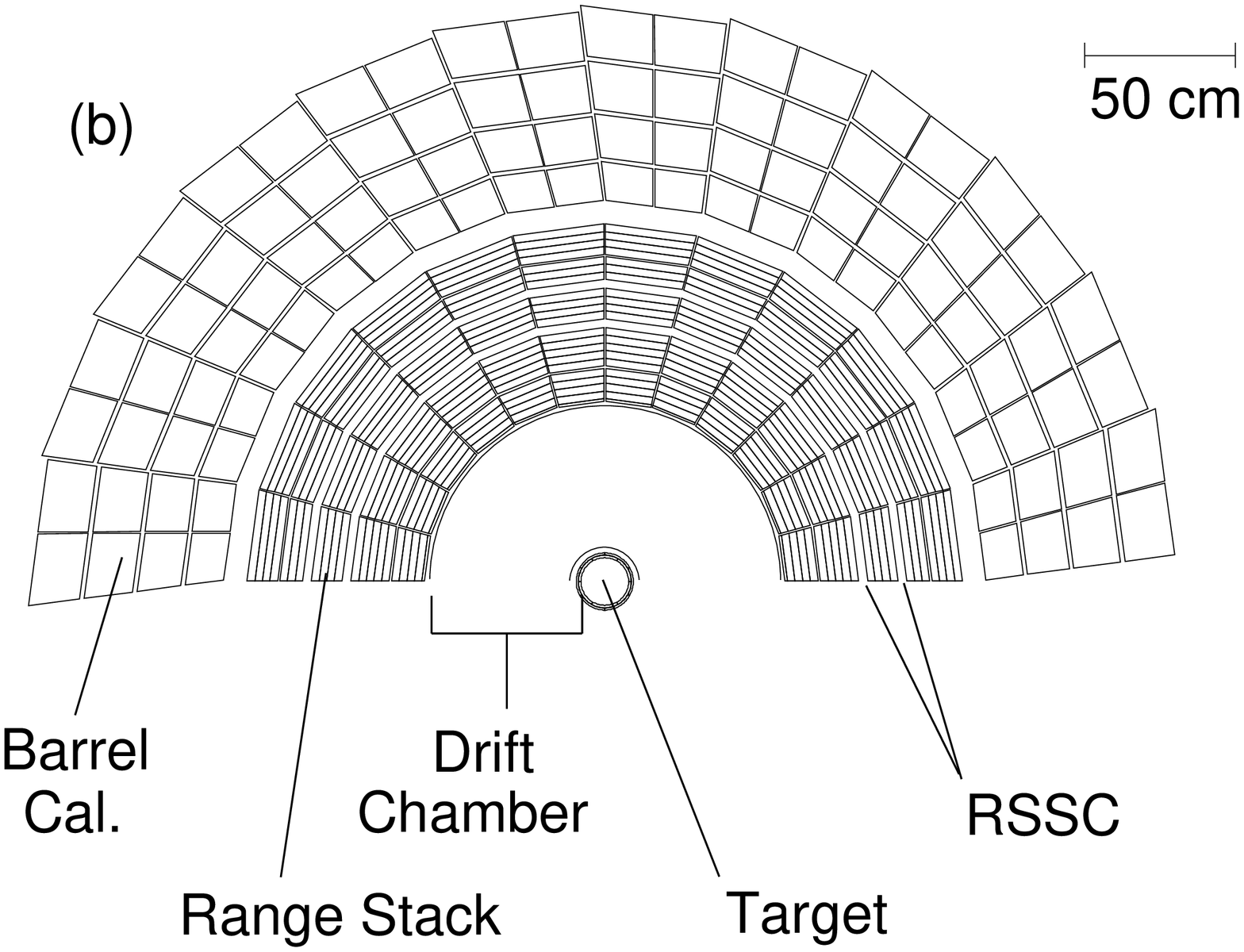,height=1.7in}
\hfill
}
\caption{\label{fig:detector}
Side-view (a) and end-view (b) of the
upper half of the E787 detector.}
\end{figure}

The E787 spectrometer (see Figure \ref{fig:detector}) was designed 
to study the decay \kpnn~
\cite{detector}.  This detector, however, is also well suited to
look for other \k\ decay modes, including \ppnn.  The
cylindrically symmetric spectrometer is immersed in a 1-Tesla magnetic
field provided by a solenoidal coil.  The innermost radial section of 
the detector is a scintillating fiber target where incoming kaons come 
to rest.  Segmentation of 5\,mm $\times$ 5\,mm provides tracking 
of the $K^+$ and decay products in the target.  Situated 
radially outward from the stopping target is a cylindrical drift chamber 
(UTC) \cite{utc}, followed by a plastic-scintillator Range Stack (RS).
The RS is segmented into 24 azimuthal sectors and 21 radial layers (a
6.35mm thick acceptance-defining layer followed by 20 19mm layers).
Embedded within the RS are two layers of straw-tube 
chambers (RSSC) operated in limited-steamer mode.
The RS photomultiplier tube (PMT) pulse shapes are 
recorded by 500-MHz flash-ADC transient digitizers (TDs)~\cite{tds}.   
In addition to providing timing and energy information, the TDs provide
pion identification by detecting the \pimue\ decay chain in the RS stopping 
counter.

The spectrometer is also equipped with an hermetic photon-detection system.  
The barrel (BL) and endcap (EC) calorimeters, encompassing nearly 
$4\pi\,$sr of solid angle, are used to detect $\gamma$ pairs to
reconstruct the $\pi^0$.

Situated immediately outside the RS, which is one radiation length ($X_0$)
thick, the BL is a 14.3-$X_0$ thick cylindrical photon detector, segmented 
into 4 radial layers and 48 azimuthal sectors, which 
accounts for two thirds of the total photon coverage.  
Extending $1.9\,$m in length, each of the 192 BL counters is made up of 
alternating layers of $1-$mm lead and $5-$mm scintillator, 
read out by PMTs at both ends.  Approximately 29\% of the energy deposited in
the BL is visible in its scintillator.  Position measurements along \(z\) 
({\it i.e.}, parallel to the beam) 
derive from both charge- and time-based information.

The two ECs on the upstream (US) and downstream (DS) ends of the 
detector account for the remaining third of the photon coverage\cite{ec}. 
Each EC consists of 4 concentric rings of undoped CsI crystals 13.5-$X_0$ 
thick.  Each EC module is viewed by a single PMT~\cite{kekpmt}, 
whose pulse shape is recorded by 500-MHz CCD-based 
TDs~\cite{ccd}.  Central holes ($10.3-$cm diameter
US and $8.4-$cm diameter DS) in the ECs accommodate the beam and 
target systems, respectively.

The photon veto encompasses a number of additional subsystems that fill 
minor openings 
along the beam direction, as well as any active parts of the detector not 
in the vicinity of the charged track.  These are used in the data analysis to 
select $\gamma$ showers that are confined to the BL and ECs and to reject 
background processes with additional photons.

The \ppnn\ trigger \cite{trig} requires a kaon to enter the 
fiber target, and an online delayed coincidence (DC) of at least 1.5\,ns is 
imposed to ensure that the kaon stops prior to its decay.  The DC is also 
necessary to suppress background from scattered beam pions \cite{pisc}.  A 
single 
charged track is required to pass through the first two RS layers, penetrate
to at least the sixth layer, and stop before the 19th layer.  To 
differentiate pion from muon tracks online, 
the $\pi^+ \rightarrow \mu^+ \nu_{\mu}$ decay signature is sought in the 
RS stopping counter by demanding that the TD
pulse of the pion have a larger area, due to the decay muon,
than a single pulse of the same height.
Tracks reaching the outer three RS layers are also vetoed, which 
effectively suppresses \kmutwo\ events.
The only photon veto imposed in the trigger is
the condition that rejects events with RS energy outside
the region of the charged track that triggers the detector.
The data set used for the signal search consists of \(1.7\times10^6\) \ppnn\ 
triggers from an exposure of about $1.3 \times 10^8$ stopped $K^+$ during 1995.

In the offline analysis, a delayed coincidence of 
2\,ns % (\(t_{track}-t_{beam}\))
is imposed to remove residual events triggered by prompt beam pions.
Events containing multiple beam particles are rejected via analysis 
of the beam counter
and chamber systems.  All events are required to pass the basic charged-track 
(CT) and \npi\ reconstruction cuts.  In the former, the CT is
required to be well reconstructed in the target, UTC and RS.
The total momentum of the CT ($P_{\pi^+}$) is determined by correcting
the momentum measured in the UTC for the energy loss suffered by a $\pi^+$
with the observed track length in the target. The resulting resolution 
is 2.74\,MeV/$c$ for  \kpitwo\,(\pitwo) events. Detailed target 
tracking of the CT identifies events in which a decay \cpi\ scatters before 
reaching the UTC.  To reject muons, a good TD double-pulse 
fit is required in the RS stopping counter, confirming the \cpi\ signature 
($\pi^+ \to \mu^+ \nu_{\mu}$).  Additional muon rejection is 
obtained by demanding the independently measured CT kinematic quantities
be consistent with those of a \cpi : (1) the UTC-measured momentum must match
the RS range and energy and (2) the RS energy-deposition pattern must be 
consistent with that of a \cpi\, (this also suppresses events with a 
photon overlapping the RS track).

Photon shower activity is identified in the various subsystems as hits in 
coincidence with the CT to within a few ns with energy above a low
threshold (typically $\sim 1\,$ MeV).   Photons are reconstructed by assembling
BL and EC hits into clusters, according to the proximity of the struck modules.
Reconstruction of the \npi\ requires the number of photon 
clusters (\Ng) to be exactly two.  Photon-veto (PV) cuts are imposed to
improve the overall \npi\ reconstruction by eliminating showers that originate
in the RS, as well as to remove additional photons from radiative decays. 

The copious decay \pitwo\ is topologically identical to \ppnn.
Thus the search for the latter must be confined to a region in which
the two processes are kinematically distinct, {\it i.e.} where the
$\pi^+$ and the $\pi^0$ are significantly softer than those from
\pitwo.  However \pitwo\ still constitutes
the dominant background, due to cases in which an inelastic 
scatter in the target down-shifts the \cpi\ energy and independently
the observed \npi\ energy fluctuates downward.   For \kmuthr\, (\muthr ) and 
\kpitwg\, (\pitwg ), the \npi\ and its charged 
partner are both naturally down-shifted by the presence of a third energetic 
particle ({\it i.e.} the neutrino in \muthr\ and the radiative photon in 
\pitwg ).  These 
backgrounds arise respectively by misidentifying the \cmu\ as a \cpi\ in 
\muthr, and missing the radiative photon in the \pitwg\ decay.
``Double-beam'' events are associated with a second beam particle
(\k\ or \cpi) that interacts or decays such that only a soft 
\cpi,\npi\ pair is seen.

Signal extraction requires that the total background be suppressed to
a small fraction of an event.  Most backgrounds are directly
determined from data; care is taken to avoid examination of potential
signal events.  
For each background study, two independent high-rejection cuts (or sets of 
cuts) are employed.  Each background type is separately enhanced by inverting
one of these cuts.  Thus the targeted background
can be isolated and any {\it other\/} cuts can be safely tuned without
fear of bias from exposure of signal events.  The roles of
the two cuts can then be reversed and the two rejections multiplied.

\begin{figure}
\hfill
\centerline{\psfig{file=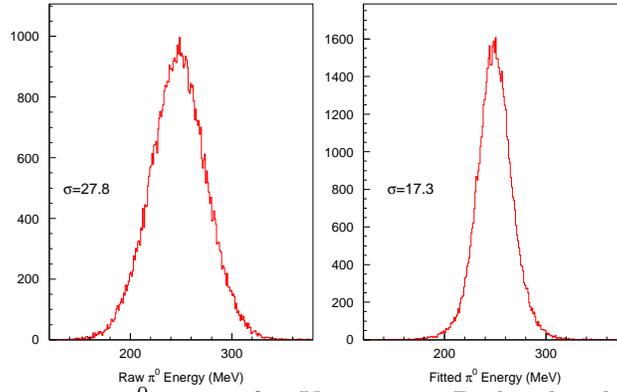,height=2.0in}}
\caption{
Left plot shows raw $\pi^0$ energy for \pitwo\ events.  Right plot
shows the $\pi^0$ energy for the same sample after the kinematic fit.}
\label{fig: reso}
\end{figure}

\begin{figure}
\hfill
\centerline{\psfig{file=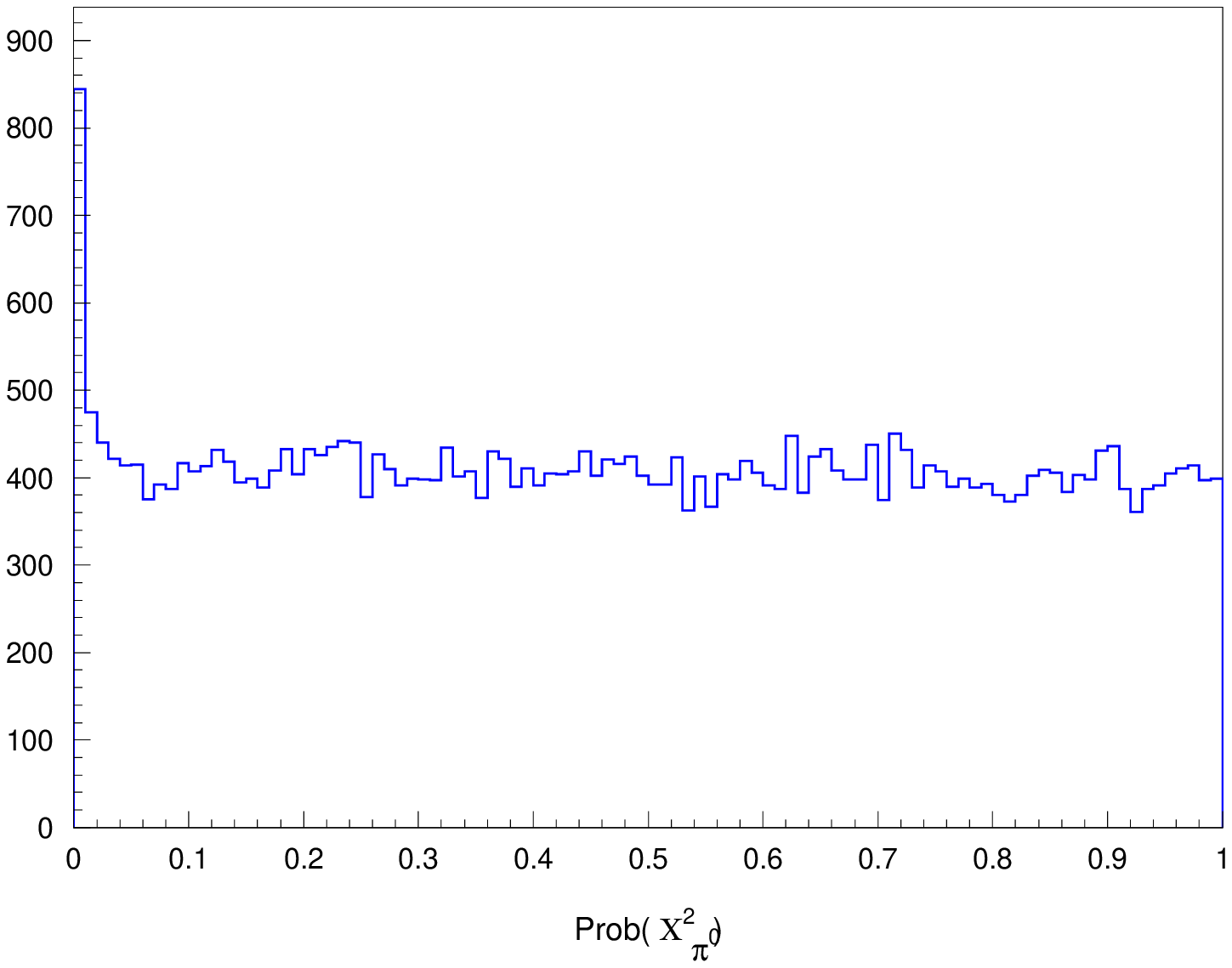,height=2in}}
\caption{Confidence level, \Probnpi\,, for \pitwo\ events.}
\label{fig: chi}
\end{figure}

To improve the rejection of most backgrounds, 
a kinematic fit to a \npi\ hypothesis is performed on events with two photons,
using energies and directions from photon clustering.  As a result,
the  resolution of the fitted \npi\ energy (\Enpi) for \pitwo\ events is
$\sigma = 17.3\,$MeV, compared with $\sigma = 27.8\,$MeV
for the raw measured \Enpi\ (see Figure~\ref{fig: reso}).   
In addition, a cut on the $\chi^2$-converted probability, \Probnpi , 
from the kinematic fit discriminates against \npi\ tail effects arising 
either from photons down-shifted in energy via shower leakage, 
or from accidental pairings in which a single \npi\ photon combines with 
a random hit or a radiative photon from \pitwg.  Such photon 
pathologies tend to have bad fits and can be removed with 
a \Probnpi\ cut at $5\%$.  The \Probnpi\ distribution for \pitwo\ events is
shown in Figure~\ref{fig: chi}.

Overlapping photon clusters responsible for certain (\Ng$>\,2$)
backgrounds ({\it e.g.}, \pitwg) typically have more hits than normal and can
often be detected by a detailed examination of the hit pattern of each 
cluster.  The \g-overlap cut set, OVLP, also exploits the fact that the
presence of two (or more) photons in a single cluster is often indicated by
a significant displacement between the two most energetic counters 
in that cluster.  These cuts are effective 
in removing coalesced showers in the BL and EC.

\begin{figure}
\centerline{\psfig{file=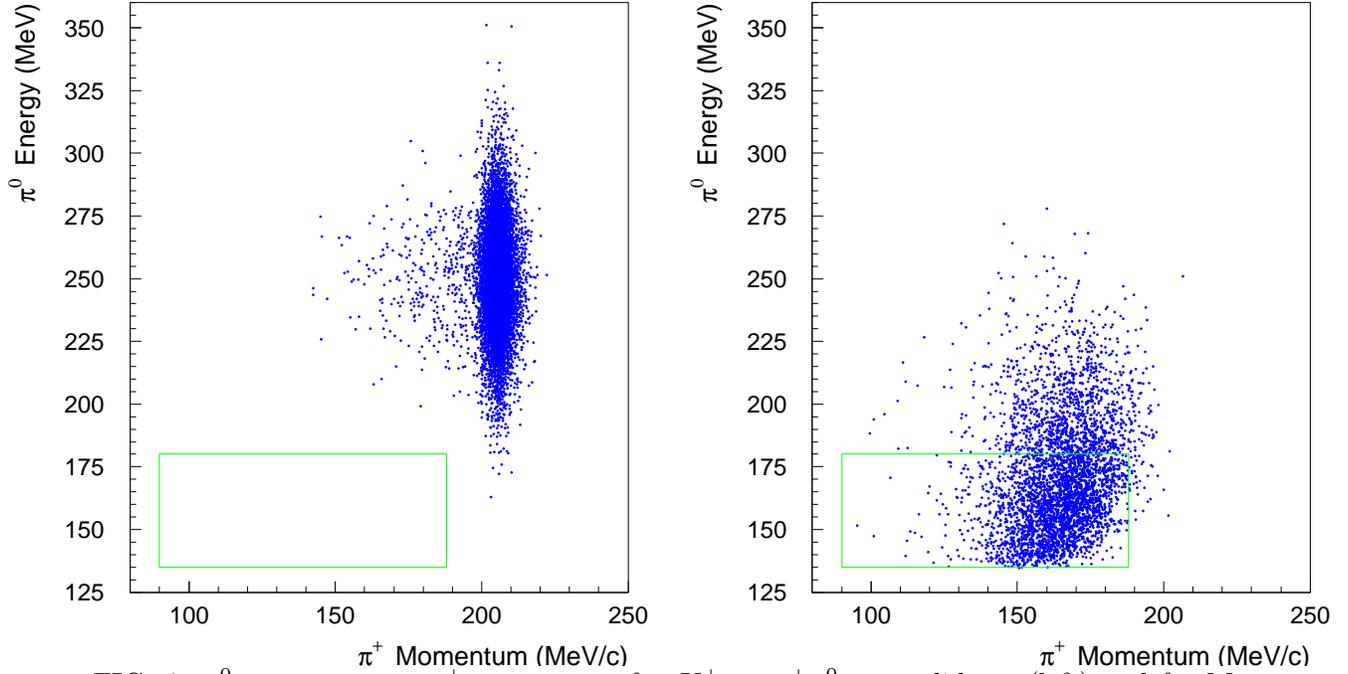}}
\caption{$\pi^0$ energy versus $\pi^+$ momentum for \kppnn\ candidates (left)
and for Monte Carlo signal events (right). Box indicates the signal
acceptance region.  \pitwo\ events cluster at the
upper right in the top plot.}
\label{fig: box} 
\end{figure}

In the case of \pitwo\ (and double-beam\footnote{In general, the 
\kpitwo\ background measurement technique is valid for all processes with 
a $\pi^+ \,, \pi^0$ pair, which includes double beam background.}) background,
advantage is taken of the fact that the \cpi\ and the \npi\ are measured by 
independent systems.  Because \pitwo\ yields mono-energetic decay 
products, to reach the signal region the \cpi\ must lose energy through 
hadronic scattering in the stopping target and at least one of the 
\npi\ photons must suffer a downward fluctuation in observed 
energy\footnote{The 
$\gamma\gamma$ opening angle must have a compensating error
to keep $m_{\gamma\gamma} \approx m_{\pi^0}$.}.
These phenomena can be partly suppressed by target and  \Probnpi\ cuts but 
these cuts alone provide only modest rejections on the \cpi,\npi\ kinematic 
tail events.  Thus cuts on \P\ and \Enpi\ are used in suppressing
the \pitwo\ background and estimating its residual. Figure~\ref{fig: box}
shows the acceptance ``box'' defined by these cuts.
The expected \pitwo\ background is found to be \(0.044\pm0.009\) events 
in the \ppnn\ search region for the present sample.

For \pitwg\ background, both data and Monte Carlo were used to
determine the number of \pitwg\ events in the \ppnn\ search region.
In rejecting \pitwg, photon multiplicity,
\Probnpi, PV, cluster-characteristics cuts ({\it e.g.}, on the number
of hits in each cluster and OVLP), and cuts on photon overlap of
the \cpi\ track are all useful.  The dominant contribution is found to
come from the inner bremsstrahlung (IB) component which 
yields \(<0.01\) events at 90\% confidence level for the present data
sample.  The contribution from the direct emission (DE) component of
\pitwg\ is negligible. For \muthr\ background, one relies on muon 
rejection via the independent TD and \cpi\ kinematic-consistency cuts. 
The expected muon background is \(0.024\pm0.012\) events.  

The total expected background from all sources is $0.068 \pm 0.021$ events.
Validation checks for the \pitwo\ and \muthr\ background estimates were
carried out by examining events outside but near the \P\ vs \Enpi\ 
box.  This requires relaxing cuts and comparing the estimated number
of residual events with the results of actual examination of the exterior 
neighborhood of the \ppnn\ box.
Good agreement is obtained between the background predictions and 
the number of events observed for several larger boxes.  The 
\pitwo\ outside-the-box background study results are
shown in Table \ref{tab: kp2bg_res_outbox} as an example.

After imposing the complete set of cuts, no events are observed in the 
search region  defined by $90\,$MeV/$c$ $<P_{\pi^+}<188\,$MeV/$c$ and 
$135\,$MeV $<E_{\pi^0}<180\,$ MeV, as shown in Figure \ref{fig: box}.  

The total acceptance for the \ppnn\ analysis is determined to be 
$0.0021 \pm 0.0001$.  Losses due to delayed coincidence,
\cpi\ and \npi\ reconstruction, timing (beam and photon vetoes), 
TD and  \cpi\ kinematic consistency cuts, and \Probnpi\ and \g-cluster
pattern-related cuts are all determined from data.  The Monte Carlo is
used only to determine acceptances associated with the trigger, 
\cpi\ nuclear-interaction and decay-in-flight losses, 
and the kinematic box cut.
These factors are listed in Table \ref{tab: acc_ppnn}.  A more 
detailed description of the analysis can be found in Reference~\cite{chi}.

Finally, the signal sensitivity is ascertained through 
the ratio of the numbers of acceptance corrected
\kppnn\ and \kpitwo\ events, both measured from the same \ppnn\
data set.  The majority of the cuts used for the \pitwo\ event selection 
are similar to those employed in the \ppnn\ analysis so that  
systematic errors incurred in the acceptance measurements tend 
to cancel.  The major differences are in the kinematic box cut 
which in this case is tuned for a \cpi\ that is monoenergetic,
an additional fiducial cut on the \cpi\ angle with respect to 
the beam direction,
and kinematic cuts exploiting the monoenergetic nature of the \npi .
The number of surviving
\pitwo\ events is 101775, corresponding to a total acceptance of 
$0.019 \pm 0.001$.  Based on zero events found in the 
\ppnn\ signal region and the \pitwo\ branching ratio, 
$0.2116 \pm 0.0014$~\cite{pdb}, we set a 90\% 
confidence level (c.l.) upper limit \bppnn\ $<\ 4.3\times10^{-5}$.

\begin{figure}
 \centerline{\psfig{file=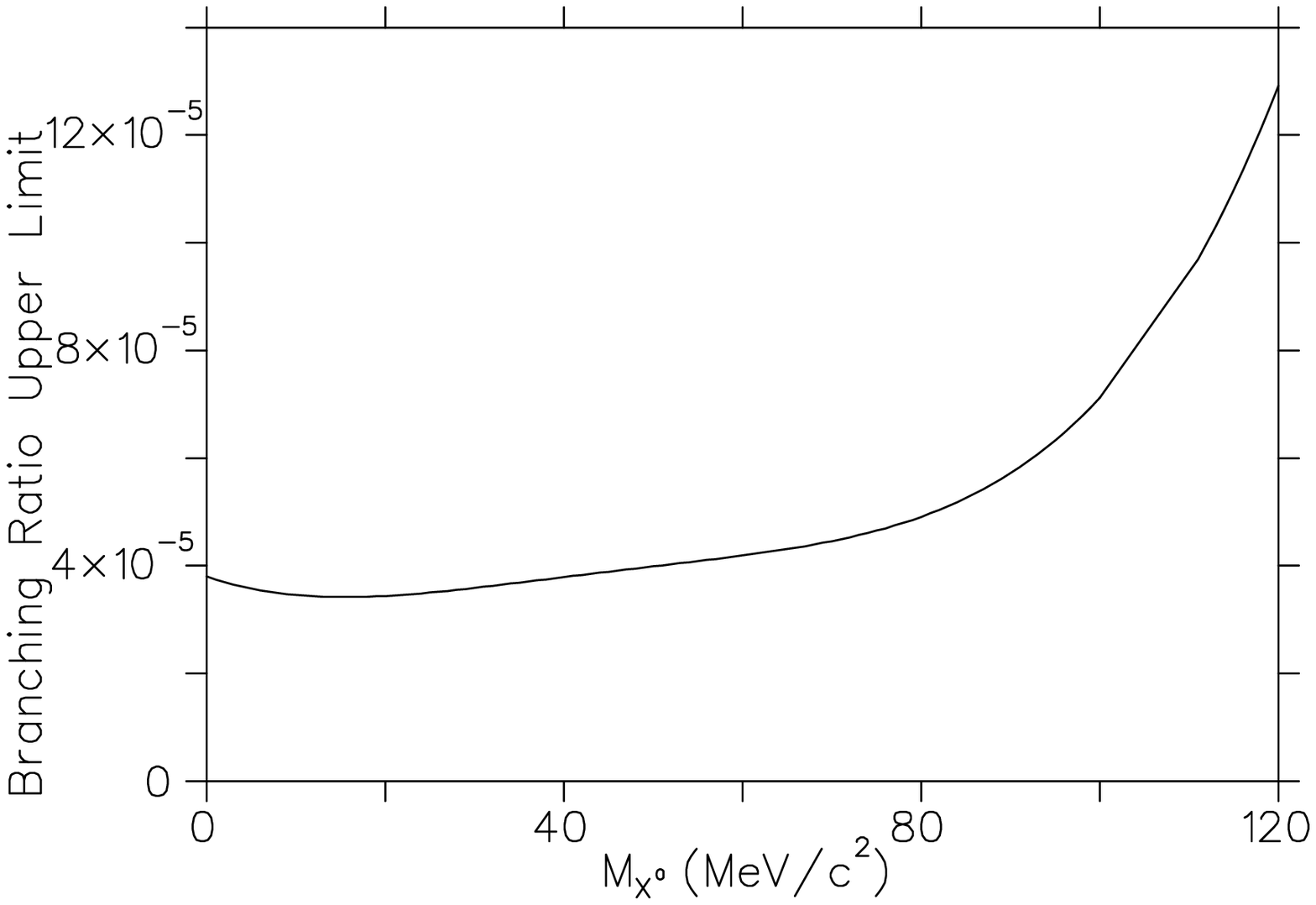,height=3in}}
 \medskip
 \caption{ The 90\% c.l. upper limit for $B(K^+ \to \pi^+ \pi^0 X^0)$
 for a three-body phase-space decay model.}
 \label{fig: X0} 
 \end{figure}

 The present data can also be used to limit the branching ratio for 
 the process $K^+ \to \pi^+ \pi^0 X^0$ where $X^0$ is a single non-interacting
 particle.  Since this is a three-body decay, the extraction of such
 limits depends on the assumed matrix element. 
 Figure~\ref{fig: X0} shows the 90\% c.l. upper limit as a
 function of $X^0$ mass under the assumption of phase space\footnote{Testing 
 a few plausible matrix elements resulted in upper limits that were 
 comparable or lower.}.

 There are good prospects to improve these limits further.  First, the total
E787 data set obtained in 1995-8 has approximately five times the sensitivity
of the search reported here.  Second, a new experiment, E949~\cite{e949},
now in preparation at BNL could yield further gains by virtue of its
much larger exposure of $K^+$ and from trigger optimization.

We gratefully acknowledge the dedicated effort of the technical staff
supporting this experiment and of the Brookhaven AGS Department.  This
research was supported in part by the U.S. Department of Energy under
Contracts No. DE-AC02-98CH10886, W-7405-ENG-36, and grant
DE-FG02-91ER40671, by the Ministry of Education, Science, Sports and
Culture of Japan through the Japan-U.S. Cooperative Research Program
in High Energy Physics and under the Grant-in-Aids for Scientific
Research, for Encouragement of Young Scientists and for JSPS Fellows,
and by the Natural Sciences and Engineering Research Council and the
National Research Council of Canada.

%-----------------------------------------------------------------------------
\pagebreak

%-------------------------------------------------------------------------------------------------------
\pagebreak

\begin{table}
\begin{center}
\begin{tabular}{|l|c|c|}
\multicolumn{3}{|c|}{Outside-the-Box \kpitwo\ Background Study}  \\ \hline
{ OUTBOX} Definition               &   Background Expectation  &   \# Events Observed  \\ \hline
  188\ $<$\ {\P\ }\ $<\ 194$  
	   				&                           &                       \\ 
  180\ $<$\ {\Enpi\ }\ $<\ 190$   
					&           0.127           &           0           \\ \hline
  188\ $<$\ {\P\ }\ $<\ 200$   
					&                           &                       \\ 
  180\ $<$\ {\Enpi\ }\ $<\ 210$   
					&           14.6            &           16          \\ \hline
  188\ $<$\ {\P\ }\ $<\ 203$   
					&                           &                       \\ 
  180\ $<$\ {\Enpi\ }\ $<\ 225$   
					&           423             &           433         \\ 
\end{tabular}
\end{center}
\caption{Outside-the-box \kpitwo\  background results (see text).}
\label{tab: kp2bg_res_outbox}
\end{table}

\begin{table}
\begin{center}
\begin{tabular}{lrr} 
Acceptance Factors   			&  \ppnn\ &\pitwo\ \\ \hline
\kppnn\ trigger acceptance     			&   
                   $\qquad\qquad\qquad\qquad 0.060$&$0.446$  \\ 
Kinematic box acceptance	&   $0.644$ & $0.966$  \\ 
% \cpi\ mass cut	        &   $0.998$& na     \\ 
\cpi\ fiducial cut                      &   na & $0.967$     \\ 
\npi\ kinematic cuts                    &   na & $0.925$     \\ 
\cpi\ nuclear interaction and decay-in-flight $^*$  &   
                                            $0.778$& $0.681$  \\ 
\cpi\ reconstruction efficiency$^*$ 	&   $0.944$ & $0.944$  \\ 
Delayed coincidence cut$^*$ 		&   $0.803$ & $0.803$  \\ 
\cpi\ target tracking, dE/dx, and timing cuts$^*$ & $0.357$ & $0.359$  \\ 
\cpi\ kinematic consistency cuts$^*$ 	&   $0.943$ & $0.970$  \\ 
Transient digitizer (\(\pi\rightarrow\mu\)) cuts$^*$ &  
                                            $0.402$ & $0.401$  \\ 
Photon cluster pattern cuts$^*$  	&   $0.946$ & $0.945$  \\ 
Requirement of two good photon clusters$^*$ &   $0.851$ & $0.850$  \\ 
 \Probnpi\ cut$^*$ 	&   $0.852$ & $0.852$  \\ \hline 
Total acceptance 			&   $0.0021$ & $0.019$ \\
\end{tabular}
\end{center}
\caption{Acceptance factors for \kppnn\,(\kpitwo). Starred entries ($*$) 
indicate 
factors that are the same or closely related in \ppnn\ and \pitwo\ 
so that errors in their determination tend to cancel in the ratio. 
`na' means not applicable}
\label{tab: acc_ppnn}
\end{table}

\begin{references}
\bibitem{trig} The trigger used for \kppnn\ was primarily devised to 
accumulate data for calibrations and acceptance measurements and was
prescaled by a factor 8656.  It was not 
optimized for \ppnn\ decay.   A requirement that the \cpi\ reach the 
sixth layer of the RS is particularly restrictive for this four-body decay.
\bibitem{pisc} In the absence of such a cut, 
beam pions which undergo hadronic scattering in the fiber target 
and are redirected toward the RS can trigger the system.
\end{references}
\end{document}